\documentclass[
12pt]{article}
\usepackage{epsf,
amsmath,amscd,amssymb,graphicx}

\begin{document}

\thispagestyle{empty}
%
\begin{flushright}
TIT/HEP--482 \\
UTP-02-45 \\
{\tt hep-th/0208127} \\
August, 2002 \\
\end{flushright}
\vspace{3mm}
\begin{center}
{\Large
{\bf Exactly solved BPS wall and winding number 
} 
\\
\vspace{2mm}{\bf in ${\cal N}=1$ Supergravity
}
} 
\\[12mm]
\vspace{5mm}

\normalsize

  {\large \bf 
  Minoru~Eto~$^{a}$}
\footnote{\it  e-mail address: 
meto@th.phys.titech.ac.jp
},  
  {\large \bf 
  Nobuhito~Maru~$^{b}$}
\footnote{\it  e-mail address: 
maru@hep-th.phys.s.u-tokyo.ac.jp
},  
 {\large \bf 
Norisuke~Sakai~$^{a}$}
\footnote{\it  e-mail address: 
nsakai@th.phys.titech.ac.jp
},
~and~~  {\large \bf 
Tsuyoshi~Sakata~$^{a}$}
\footnote{\it  e-mail address: 
sakata@th.phys.titech.ac.jp
} 

\vskip 1.5em

{ \it $^{a}$Department of Physics, Tokyo Institute of 
Technology \\
Tokyo 152-8551, JAPAN  \\
and \\
  $^{b}$Department of Physics, University of Tokyo 
113-0033, JAPAN
 }
\vspace{5mm}
{\bf Abstract}\\[5mm]
{\parbox{13cm}{\hspace{5mm}
A BPS exact wall solution is found in ${\cal N}=1$ supergravity 
in four dimensions. 
The model uses chiral scalar field with a periodic 
superpotential admitting winding numbers.  
Maintaining the periodicity in supergravity 
requires a gravitational correction to 
superpotential which allows the exact solution. 
By introducing boundary cosmological constants, 
we construct non-BPS multi-wall solutions for which 
a systematic analytic approximation is worked out 
for small gravitational coupling. 
}}
\end{center}
\vfill
\newpage
\setcounter{page}{1}
\setcounter{footnote}{0}
\renewcommand{\thefootnote}{\arabic{footnote}}

{\large\bf Introduction}

\vspace{2mm}

Supersymmetry (SUSY) has been most promising for unified theories beyond 
the standard model \cite{DGSW}. 
Recently new possibilities have been added by 
the  ``Brane World" scenario, where 
 our four-dimensional spacetime 
is realized on topological defects such as walls 
 embedded in a higher dimensional spacetime \cite{LED}. 
More intriguing model has also been proposed by 
using  the so-called warped metric 
 in a bulk AdS spacetime of higher dimensional gravity 
together with an orbifold 
$S^1/Z_2$ with a boundary cosmological constant 
finely tuned with the bulk cosmological constant \cite{RS}. 
The warped spacetime has provided an attractive notion of localized 
graviton on the wall \cite{RS2}. 
Much efforts have been devoted to implement supergravity 
in orbifolds \cite{ABN}, 
work out solutions in gravity with 
a scalar field to replace the orbifold 
by a smooth wall configuration in four dimensions 
\cite{CGR, BCY} 
and in higher dimensions \cite{BCY}--\cite{Sasakura}. 
It has been noted that the set of second order differential equations 
for coupled scalar and gravity theories can be reduced to a set of 
first order differential equations using a potential similar to 
the superpotential even for nonsupersymmetric case \cite{DFGK}. 

In SUSY theories, single wall can preserve half of the 
SUSY, and is called a BPS state \cite{WittenOlive}. 
Properties of topological defects such as walls 
have been extensively studied in SUSY theories 
\cite{DvSh}-\cite{SaSu} and exact solutions have been useful as 
illustrated by a $1/4$-BPS junction of walls \cite{OINS}. 
Coexistence of BPS walls conserving different supercharges can break 
SUSY completely \cite{MSSS}. 
When SUSY is broken, the stability of the multi-wall system 
is no longer guaranteed. 
We have found that topological quantum number such as winding number can 
stabilize the non-BPS multi-wall configurations. 
In an ${\cal N}=1$ globally SUSY 
sine-Gordon model, 
we have constructed an exact non-BPS multi-wall solution 
with a nonvanishing winding number which has no tachyon and 
is stable 
in a compact base space with the radius $R$ \cite{MSSS2}. 
We have also found 
an example of a stable non-BPS bound state of BPS walls 
 in an extended model \cite{SaSu}. 
Since the compact space radius $R$ becomes a dynamical quantity 
in gravity theories, we are led to ask if the stabilization 
mechanism due to the winding number still works in supergravity. 
To study properties of walls in supergravity theories, we first 
need to find solutions of BPS and non-BPS walls. 

The purpose of our paper is to present an exact BPS solution of 
the SUSY sine-Gordon model embedded into ${\cal N}=1$ supergravity in four 
dimensions and a systematic analytic approximation for non-BPS multi-wall 
configurations in powers of gravitational coupling. 
We find that a gravitational correction to the 
superpotential is needed to maintain the periodicity of our 
SUSY sine-Gordon model as an isometry realized 
by a K\"ahler transformation. 
This allows us to obtain an exact BPS wall solution in 
the ${\cal N}=1$ supergravity. 
Since equations of motion for matter and gravity are simultaneously 
satisfied, 
our solution requires no artificial fine-tuning of parameters 
such as boundary and bulk cosmological constants contrary to the original 
orbifold model \cite{RS}. 
We discuss classification and global properties of non-BPS 
solutions using the first order differential equations of 
Ref.\cite{DFGK}, and 
 succeed in constructing multi-wall configurations 
with and without winding number using scalar field. 
We show that multi-wall configurations 
require negative energy density. 
We introduce a negative energy density as a boundary 
cosmological constant similarly to \cite{RS, RS2}. 
The scalar field satisfies the equation of motion without any 
external sources and is smooth everywhere. 
We also obtain a systematic analytic approximation in powers of 
gravitational coupling and express necessary boundary cosmological 
constant in terms of elliptic integrals, for instance. 
To this end, we identify a set of first order differential equations 
for walls with single scalar field 
corresponding to the no gravity limit of the method of \cite{DFGK}. 
We analyze the region of large gravitational coupling 
numerically and find qualitatively similar results.

\vspace{4mm}

{\large\bf Exact BPS Wall Solutions with a Periodic Potential
}

\vspace{2mm}

To avoid inessential 
 complications, we will consider three-dimensional walls 
 in four-dimensional theory to obtain 
BPS and non-BPS solutions except stated otherwise. 
Previously we used a simple model admitting the winding number 
in ${\cal N}=1$ SUSY theory, 
the SUSY sine-Gordon model \cite{MSSS} 
with the K\"ahler potential $K$ 
for the minimal kinetic term 
and the periodic superpotential $P$ for a chiral scalar superfield 
$\Phi=(\phi, \psi, F)$ 
\begin{equation}
K(\Phi^{\dagger}, \Phi) = \Phi^{\dagger}\Phi, \qquad 
P_{\rm global}(\Phi)
=\frac{\Lambda^{3}}{g^{2}}\sin\left(\frac{g}{\Lambda}\Phi\right) .
\end{equation}
The model is invariant under 
\begin{equation}
\label{eq:periodicity}
\phi \rightarrow \phi + {2n\pi \Lambda \over g}, \qquad n \in {\bf Z}
\end{equation}
Coupling the model with the minimal kinetic term 
to ${\cal N}=1$ supergravity \cite{WessBagger} 
 gives a 
Lagrangian whose bosonic part is given in terms of the 
superpotential $P(\phi)$ with scalar field $\phi$ replacing 
the superfield $\Phi$ 
\begin{equation}
\label{eq:sugra_lagrangian}
(\sqrt{- g})^{-1}\mathcal{L} =
-\dfrac{M_{\rm P}^2}{2}R 
-g^{mn}\partial_m\phi\partial_n\phi^*
-{\rm e}^{\frac{|\phi|^2}{M_P^2}}
\left[\left|{\rm e}^{-{|\phi|^2} \over M_{\rm P}^2}
{\partial \over \partial \phi}\left({\rm e}^{{|\phi|^2} \over M_{\rm P}^2}
 P\right)\right|^2-{3 \over M_{\rm P}^2}|P|^2\right] ,
\end{equation}
where $g_{mn}$ is a metric of the spacetime, $g={\rm det}g_{mn}$, 
$R$ is the scalar 
curvature, and $M_{\rm P}$ 
is the Planck mass\footnote
{We follow the convention of Ref.\cite{WessBagger} except that the 
Einstein indices are denoted by $m, n, \dots$ in four dimensions, 
and by $\mu, \nu, \dots$ in three dimensions. 
} 
.

Since we wish to study the winding number of $\phi$, 
the periodicity under (\ref{eq:periodicity}) 
should be maintained. 
The K\"ahler potential transforms under (\ref{eq:periodicity}) 
as 
\begin{equation}
K(\Phi^{\dagger}, \Phi) \rightarrow 
K(\Phi^{\dagger}, \Phi) 
+ {2n\pi \Lambda \over g}\left(\Phi^{\dagger}+\Phi\right) 
 + \left({2n\pi \Lambda \over g}\right)^2 . 
\end{equation}
The Lagrangian is invariant under such K\"ahler transformations 
in the globally SUSY theory. 
In supergravity, however, 
it has been known that 
 the superpotential should transform in a definite way 
under K\"ahler transformations to make the theory invariant \cite{WessBagger}. 
We find the necessary gravitational corrections to the superpotential 
which realizes the desired periodicity (\ref{eq:periodicity}) as 
\begin{equation}
K(\Phi^{\dagger}, \Phi) = \Phi^{\dagger}\Phi, \qquad 
P(\Phi)
={\rm e}^{-{\Phi^2 \over 2M_{\rm P}^2}}P_{\rm global}(\Phi)
={\rm e}^{-{\Phi^2 \over 2M_{\rm P}^2}}
\frac{\Lambda^{3}}{g^{2}}\sin\left(\frac{g}{\Lambda}\Phi\right) .
\end{equation}
This gravitational correction is crucial to obtain 
exact periodicity and also an exact solution 
of BPS walls. 

We find that the 
SUSY vacuum condition in supergravity gives precisely the same 
SUSY vacua as in the globally SUSY case, 
$\tilde\phi\equiv g\phi/\Lambda={\pi \over 2}+n\pi, \; n \in {\bf Z}$ 
\begin{equation}
0 = {\rm e}^{-{|\phi|^2 \over M_{\rm P}^2}}{\partial \over \partial\phi} 
\left[{\rm e}^{|\phi|^2 \over M_{\rm P}^2} P(\phi)\right]
=
{\rm e}^{-\phi^2 \over 2M_{\rm P}^2}{\Lambda^2 \over g} 
\left[\cos {g \over \Lambda}\phi + 
{\Lambda(\phi^* - \phi) \over M_{\rm P}^2}
{1 \over g}\sin {g \over \Lambda}\phi \right] .
\label{eq:cond_SUSYvacu}
\end{equation}
No other vacua are allowed in spite of nontrivial dependence 
on $M_{\rm P}$. 
By requiring half of the SUSY to be conserved, 
one obtains the BPS equations in ${\cal N}=1$ supergravity. 
We assume the following warped metric Ansatz \cite{RS} 
\begin{equation}
ds^2 =  g_{mn}dx^mdx^n = e^{2A(y)}\eta_{\mu\nu}dx^\mu dx^\nu + dy^2 ,\ \ 
(\mu,\nu=0,1,3)
\label{eq:warped_metric}
\end{equation}
where $A(y)$ is the 
warp factor,  $\eta_{\mu\nu} = \text{diag}.(-,+,+)$, and 
the extra dimension is denoted as $x^2=y$.
Parametrizing the conserved SUSY parameter as  
$\zeta(y)={\rm e}^{i\theta(y)}\sigma^2\bar{\zeta}(y)$, 
the half SUSY condition for gravitino gives the BPS equation for 
the Killing spinor $\zeta_\alpha$ and the warp factor $A$  \cite{CGR}
\begin{eqnarray}
\label{eq:BPS_Killing_sp}
{d \theta \over dy} &\!\!\!=&\!\!\! 
{i \over 2M_{\rm P}^2}\left(\phi^*{d \phi \over dy}
-\phi{d\phi^* \over dy}\right) , \qquad 
{d |\zeta_\alpha| \over dy}
=
{1 \over 2}{dA \over dy}|\zeta_\alpha| , \\
\label{eq:BPS_warp_ft}
{d A \over dy} &\!\!\!=&\!\!\! 
-i{\rm e}^{i\theta} 
\dfrac{-1}{M_{\rm P}^2}{\rm e}^{\frac{|\phi|^2}{2M_{\rm P}^2}}P(\phi) .
\end{eqnarray}
The BPS equation for the scalar field reads 
\begin{equation}
\label{eq:BPS_scalar}
{d \phi \over dy} 
\!=\! 
-i{\rm e}^{i\theta} 
{\rm e}^{\frac{-|\phi|^2}{2M_{\rm P}^2}}
{\partial \over \partial\phi^*}
\left({\rm e}^{\frac{|\phi|^2}{M_{\rm P}^2}}P^*\right)
\!=\! 
-i{\rm e}^{i\theta} 
{\rm e}^{\phi^*(\phi-\phi^*) \over 2M_{\rm P}^2}
{\Lambda^2 \over g}\left[\cos {g \over \Lambda}\phi^* 
+{\Lambda(\phi-\phi^*) \over g M_{\rm P}^2} 
\sin {g \over \Lambda} \phi^*\right]. 
\end{equation}

\begin{figure}[h]
\begin{center}
\includegraphics[width=10cm,height=3cm]
{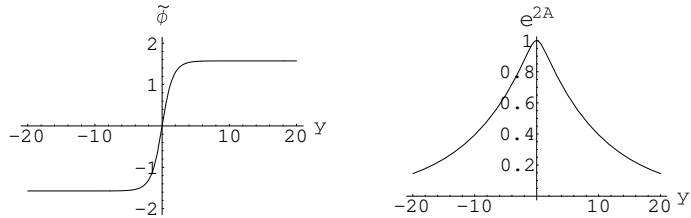}
\caption{\small BPS wall solution $\tilde \phi(\tilde y)$ 
and $A(\tilde y)$ 
 connecting $\tilde \phi =-\pi/2$ to $\tilde \phi=\pi/2$. 
}
 \label{fig:BPS_wall}
\end{center}
\end{figure}

Let us solve these BPS equations from $y=-\infty$ 
choosing a SUSY vacuum 
$\tilde\phi=-\pi/2$ and $\theta=\pi/2$ as the initial 
condition at $y=-\infty$. 
We observe that the right-hand side of the 
BPS equation for the imaginary part $(\phi-\phi^*)/2i$ 
is proportional to the imaginary part itself if $\theta=\pi/2$. 
Then the right-hand side of the 
BPS equation (\ref{eq:BPS_scalar}) for the imaginary part 
$(\phi-\phi^*)/2i$ vanishes at $y=-\infty$. 
Similarly the right-hand side of (\ref{eq:BPS_warp_ft}) 
for the phase $\theta(y)$ vanishes at $y=-\infty$. 
Therefore these first order differential equations dictate 
 vanishing imaginary part of $\phi$ and 
the constant phase of $\zeta_\alpha$ for any $y$ : 
$\theta(y)=\pi/2, \ \phi(y)=\phi(y)^*$. 
Then equation (\ref{eq:BPS_Killing_sp}) 
can be integrated to give the Killing spinor 
$
\zeta_\alpha = -\zeta_\alpha^* = ie^{\frac{1}{2}A} .
$
The remaining  BPS equation 
for $\phi=\phi^*$ becomes identical to the globally SUSY case 
and that for $A$ becomes 
\begin{eqnarray}
{d \phi \over dy} 
=
{\Lambda^2 \over g}\cos {g \over \Lambda}\phi , 
\qquad 
{d A \over dy} 
=
 -\dfrac{\Lambda^3}{g^2M_{\rm P}^2}
\sin\dfrac{g}{\Lambda}\phi .
\end{eqnarray}
Anti-BPS equation conserving the opposite SUSY 
charges is given by $\theta=-\pi/2$. 
We find solutions for a wall localized at $y_0$ connecting 
from $\tilde\phi=-\pi/2+n\pi$ at $y=-\infty$ to $\tilde\phi=\pi/2+n\pi$ 
at $y=\infty$ for any $n\in{\bf Z}$ as (anti-)BPS solution for $n=$ 
even (odd) as shown in Fig.\ref{fig:BPS_wall} 
\begin{eqnarray}
\phi
\!=\!
\dfrac{\Lambda}{g}\left[(-1)^n \left(2\tan^{-1}
{\rm e}^{(-1)^n \Lambda(y-y_0)}-{\pi \over 2}\right)
\!
+n\pi\right],
\quad 
{\rm e}^{2A(y)}
\!=\!
\left[\cosh \Lambda(y-y_0)\right]^{-
{1 \over g^2}\left({\Lambda \over M_{\rm P}}\right)^{2}}
\label{eq:BPS_sol}
\end{eqnarray}
where we suppressed an integration constant for $A$ which amounts to 
an irrelevant normalization constant of metric. 
The Killing spinor is given with the normalization 
$N$  
$\zeta_\alpha= 
iN\left[\cosh \Lambda(y-y_0)\right]^{-{\Lambda^2/(2g^2M_{\rm P}^2)}}$. 
We have verified that these BPS solutions indeed satisfy 
the equations of motion. 
The energy density of the scalar field of the (anti-)BPS wall 
is given by $4T^3\equiv 4\Lambda^3/g^2$. 
\vspace{-5mm}

Let us examine the thin-wall limit \cite{BCY}. 
By fixing the energy density $4T^3$ 
and taking a limit $g\rightarrow \infty$, 
we recover the  Randall-Sundrum model \cite{RS2}. 
The kinetic energy of the scalar field $|d\phi/dy|^2$ 
reduces to the half of the boundary cosmological constant 
$\lambda_{\text{boundary}} 
= 4T^3= 4\Lambda^3/g^2$, and the remaining half comes from the 
first term of the scalar potential in Eq.(\ref{eq:sugra_lagrangian}) 
containing derivative of the superpotential. 
The bulk cosmological constant 
$\lambda_{\text{bulk}}=-3T^6/M_{\rm P}^2$ comes from the last term 
$-3{\rm e}^{|\phi|^2/M_{\rm P}^2}|P|^2/M_{\rm P}^2$ of the scalar potential 
 in Eq.(\ref{eq:sugra_lagrangian}). 
These cosmological constants automatically satisfy the relation 
$\lambda_{\text{boundary}} 
= 4M_{\rm P}\sqrt{-\lambda_{\text{bulk}}/3}$ 
which was implemented as a fine-tuning in the orbifold model 
\cite{RS}. 
Our solutions (\ref{eq:BPS_sol}) replace 
the boundary cosmological 
constant at a fixed point of the orbifold by a smooth physical 
wall configuration of the scalar field completely. 
The existence of the thin wall limit suggests that the 
${\cal N}=1$ supergravity in four dimensions is consistent 
even on the orbifold, although we do not know if this point 
has been demonstrated similarly to the five-dimensional case 
\cite{ABN}. 

We can use the ``effective supergravity'' to extend our 
model to arbitrary spacetime dimensions $D$ \cite{SkTo}.  
Assuming the periodic potential 
$W(\phi)=(\Lambda^{D-1}/g^2) \sin (g\phi/\Lambda^{(D-2)/2})$ 
can be used, 
we obtain the $D$ dimensional BPS wall solution 
 connecting from $\tilde \phi =-\pi/2$ to $\tilde \phi=\pi/2$ 
\begin{equation}
\phi 
=
 {\Lambda^{D-2 \over 2} \over g}
\left(2\tan^{-1}e^{\Lambda(y-y_0)}-\dfrac{\pi}{2}\right),
\quad 
{\rm e}^{2A(y)}=\left[\cosh \Lambda(y-y_0)\right]^{-
{4 \over (D-2)g^2}\left({\Lambda \over M_D}\right)^{D-2}}, 
\end{equation}
with $M_D$ as the Planck mass in $D$ dimensions. 
A BPS wall different from ours 
has been obtained in a similar model 
in five-dimensional supergravity \cite{BeDA}. 

\vspace{4mm}

{\large\bf Classification and global properties of Non-BPS Solutions  
}

\vspace{2mm}

In order to study possible multi-wall configurations, 
especially non-BPS configurations, we need to solve 
the equations of motion which are second order differential equations. 
It has been shown that these second order equations are equivalent 
to a set of first order differential equations which resemble 
the BPS equations provided there is only single scalar field 
forming walls \cite{DFGK}. 

Let us assume that initial conditions for complex scalar 
$\phi(y), d\phi(y)/dy$ 
and the warp factor $A(y), dA(y)/dy$ are real at some $y$, say $y=-\infty$. 
The equations of motion for $\phi$ and $A$ in our model then 
dictate that both $\phi$ and $A$ are real for any $y$. 
Therefore we can ignore the imaginary part of the complex 
scalar field as long as we are interested in classical solutions. 
The bosonic part of the Lagrangian is given by 
\begin{equation}
\label{eq:scalar_gravity}
(\sqrt{- g})^{-1}\mathcal{L} 
= -\dfrac{M_{\rm P}^2}{2}R 
-\! \left({d \phi \over dy}\right)^2 \!\!- V(\phi), 
\quad 
V(\phi) = \dfrac{\Lambda^4}{g^2} \! \left(\cos^2\dfrac{g}{\Lambda}\phi 
- \dfrac{3\Lambda^2}{g^2M_{\rm P}^2}\sin^2\dfrac{g}{\Lambda}\phi\right).
\end{equation}
We can now apply the method of Ref.\cite{DFGK} 
to our model taking $\phi$ as a real scalar field. 
Given the scalar potential (\ref{eq:scalar_gravity}), 
we should find a real function 
$W(\phi)$ by solving the following first order nonlinear differential 
equaiton 
\begin{equation}
\label{eq:W_phi_diff_eq}
V(\phi)=\left({d W(\phi) \over d\phi}\right)^2
-{3 \over M_{\rm P}^2}W^2(\phi) .
\end{equation}
Then $\phi(y)$ and $A(y)$ are obtained by solving the 
following two first order differential equations 
\begin{equation}
{d \phi(y) \over dy}=\dfrac{d W(\phi)}{d \phi}, \qquad 
{d A(y) \over dy} = - \dfrac{1}{M_{\rm P}^2}W(\phi) .\label{phi-A}
\end{equation}
It has been shown that these equations are equivalent to the equations 
of motion \cite{DFGK}. 

It is convenient to rewrite these equations in terms of 
dimensionless variables  
\begin{equation}
\tilde{\phi} \equiv \dfrac{g}{\Lambda}\phi,
\quad
\tilde{W}\equiv {g^2 \over \Lambda^3}W,
\quad 
\tilde{V}\equiv {g^{2} \over \Lambda^4}V,
\quad 
\tilde{y}\equiv \Lambda y,
\quad 
\alpha\equiv {\Lambda \over g M_{\rm P}}, 
\label{eq:resc_W_phi}
\end{equation}
\begin{equation}
{d \tilde{W} \over d\tilde \phi}
= 
\pm \sqrt{\tilde{V}(\tilde{\phi})
+3\alpha^2\tilde{W}(\tilde{\phi})^2},
\qquad 
\tilde{V}(\tilde{\phi}) 
= 
\cos^2\tilde{\phi}
-3\alpha^2\sin^2\tilde{\phi}
\label{1st-W}
\end{equation}
\begin{equation}
\dfrac{d\tilde{\phi}}{d\tilde{y}} 
=
 \dfrac{d \tilde{W}(\tilde{\phi})}{d\tilde{\phi}}, 
\qquad 
\dfrac{dA}{d\tilde{y}} 
=
 -\alpha^2\tilde{W}(\tilde{\phi}).
\label{eq:DFGK_eq}
\end{equation}
\begin{figure}[h]
\begin{center}
\includegraphics[width=10cm,height=5cm]
{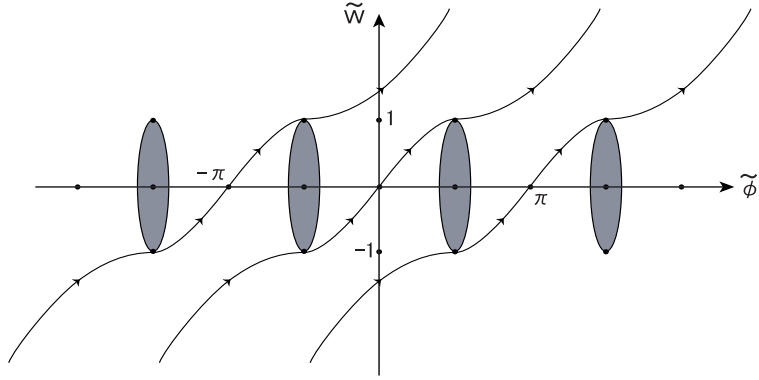}
 \caption{\small Solution curves of Eq.(\ref{1st-W}) 
with $d\tilde W/d\tilde \phi \ge 0$ 
 in $\tilde \phi \tilde W$ plane. 
 Forbidden region is shaded. 
 }
 \label{fig:W_phi_flow}
\end{center}
\end{figure}
\vspace{-5mm}

To obtain a real solution, the right-hand side of 
Eq.(\ref{1st-W}) 
has to be real : 
$\tilde V(\tilde \phi)+3\alpha^2\tilde{W}^2 \ge 0$. 
The shaded region in Fig.\ref{fig:W_phi_flow} 
shows the forbidden region 
in $\tilde \phi$ $\tilde W$ plane.
It is easy to see that the exact (anti-)BPS solutions are those lines 
whose starting and ending points are both 
tangent at the top or bottom of these forbidden regions 
as shown in Fig.\ref{fig:W_phi_flow}. 
Any solution is represented by a 
curve in $\tilde \phi$ $\tilde W$ plane. 
Since the first order differential equation (\ref{1st-W}) determines 
the solution uniquely once a value of $\tilde W$ is specified at some 
$\tilde \phi$ as an initial condition, 
we observe: 
1) No solution curves can intersect.  
Entire allowed region is covered by the curves once. 
2) Curves can end only at the boundary of forbidden region 
or at infinity. 

Taking the positive sign for $d\tilde W/d\tilde \phi$ in 
Eq.(\ref{1st-W}), the asymptotic behavior of 
solution curves are given by 
$\tilde W \sim \pm \exp 
(\pm \sqrt3 \alpha \tilde \phi) 
\rightarrow \pm \infty$. 
As illustrated 
in Fig.\ref{fig:W_phi_flow}, 
there are two different types of curves 
except the BPS solution curve : 
\begin{enumerate}
\item
Starting at the boundary of forbidden region and ending at 
$\tilde \phi, \tilde W \rightarrow \infty$.  
\item
Ending at  the boundary of forbidden region and starting at 
$\tilde \phi, \tilde W \rightarrow -\infty$.  
\end{enumerate}
Since the asymptotic behavior $\tilde \phi, \tilde W \rightarrow \pm \infty$ 
gives singularities \cite{DFGK}, \cite{Sasakura} at finite $y$ 
($y \sim y_{\pm \infty} \mp {\rm e}^{\mp \tilde \phi}/(3\alpha^2)$ as 
$\tilde \phi \rightarrow \pm \infty$), 
we conclude that BPS wall solution 
(\ref{eq:BPS_sol}) 
is the only solution which are regular in the entire region of 
$y$ if we do not 
allow any external sources for scalar $\phi$ as well as gravity, such 
as a boundary cosmological constant. 
If one looks into $\tilde \phi, \tilde W$ plane only, 
$\tilde W = \sin \tilde \phi$ may appear a multi-wall winding solution. 
However, wall solution should be considered as a function of our base 
space $-\infty< y < \infty$. 
The solution curve approaches to the top or bottom of the 
forbidden region only at $y \rightarrow \pm \infty$. 
Therefore only a segment of $\tilde W = \sin \tilde \phi$ 
between the top and the bottom of the forbidden region 
constitutes a wall solution which is a single wall. 
The non-existence of multi-wall BPS solutions is perhaps a peculiarity of 
our model which is in interesting contrast to the monopole or 
instanton case. 

We now wish to find a non-BPS solution by allowing a source due to 
a boundary cosmological constant $\lambda_i\delta(y-y_i)$ 
at appropriate points $y=y_i$ 
in the extra dimension  into 
the Lagrangian (\ref{eq:scalar_gravity}) 
similarly to the orbifold model \cite{RS, RS2}. 
We require that the equation of motion for the scalar field is satisfied 
without any external sources. 
Then only the equation of motion for the warp factor is 
modified to give the boundary condition 
\begin{equation}
{dA \over dy}(y_i+\varepsilon) -
{dA \over dy}(y_i-\varepsilon) 
= -\dfrac{1}{2M_{\rm P}^2}\lambda_i .
\label{eq:bound_cod_A}
\end{equation}
{}The boundary condition combined with Eq.(\ref{phi-A}) 
relates the boundary cosmological constant $\lambda_i$ 
 at the connection point $\phi=\phi_i$ to a discontinuity 
of $W$ at $\phi=\phi_i$ 
\begin{equation}
\lambda_i
=2 \left(W (\phi_i+\epsilon)-W (\phi_i-\epsilon)\right) .
\label{eq:cosm_const_W}
\end{equation}
Equation of motion for $\phi$ without external sources 
requires both $\tilde \phi$ and 
$d\tilde\phi/d\tilde y=d\tilde W/d\tilde \phi$ 
to be continuous across $y_i$. 

\begin{figure}[h]
\begin{center}
\includegraphics[width=15cm,height=4cm]
{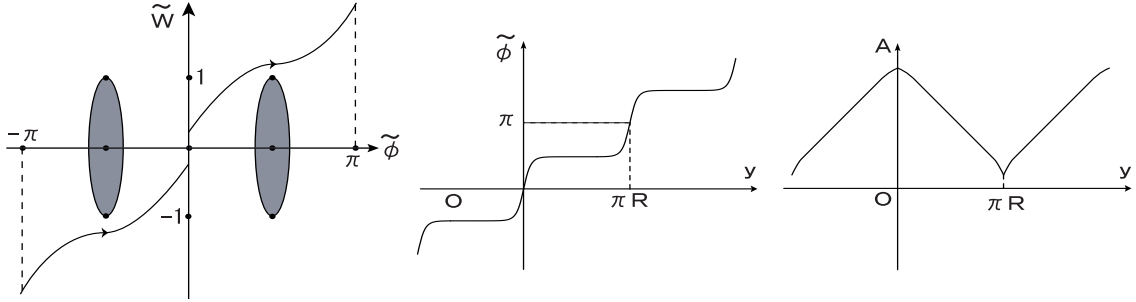}
 \caption{\small 
 Non-BPS solution with unit winding number. 
  (a)  $\tilde \phi \tilde W$ 
 plane,  (b) $\tilde \phi(\tilde y)$, and 
 (c)  $A(\tilde y)$. 
 Discontinuities in $\tilde W(\tilde \phi)$ at 
 $\tilde \phi=0, \pm\pi$ correspond to boundary cosmological 
 constants. 
 }
 \label{fig:wind_sol}
\end{center}
\end{figure}

Let us denote the solution of Eq.(\ref{1st-W}) 
specified by the initial 
condition $(\tilde \phi_0, \tilde W_0)$ as 
$\tilde W_{\pm}(\tilde \phi; (\tilde \phi_0, \tilde W_0))$ 
with the suffix $+(-)$ corresponding to the sign 
of $d\tilde W/d\tilde\phi$. 
Symmetry of Eq.(\ref{1st-W}) gives 
\begin{equation}
\tilde W_{\pm}(\tilde \phi; (\tilde \phi_0, \tilde W_0))
=
\tilde W_{\pm}(\tilde \phi+\pi; (\tilde \phi_0+\pi, \tilde W_0))
=
-\tilde W_{\pm}(-\tilde \phi; (-\tilde \phi_0, -\tilde W_0))
,
\label{eq:symm_sol_curve}
\end{equation}
\begin{equation}
\tilde W_{+}(\tilde \phi; (\tilde \phi_0, \tilde W_0))
=
\tilde W_{-}(-\tilde \phi; (-\tilde \phi_0, \tilde W_0))
\label{eq:symm_sol_curve2}
\end{equation}
As a simple solution to satisfy the requirement of continuity 
of  $\tilde \phi$ and 
$d\tilde\phi/d\tilde y=d\tilde W/d\tilde \phi$, 
we can choose to switch solution curves at 
$\tilde \phi=\pi$ 
from 
$\tilde W(\tilde \phi;(0, \tilde W_0>0))$ to 
 $\tilde W(\tilde \phi; (2\pi, -\tilde W_0))$ 
and then at $\tilde \phi=2\pi$ 
to $\tilde W(\tilde \phi; (2\pi, \tilde W_0))$. 
As shown in Fig.\ref{fig:wind_sol} this solution 
is periodic in $\tilde \phi$ with the periodicity 
$2\pi$, and has a unit winding number and 
the symmetry $S^1/Z_2$ which is 
consistent with the orbifold compactification as in 
Refs.\cite{RS}, \cite{RS2}. 
We see that we cannot avoid negative cosmological constant 
(negative energy density). 
{}For small $\tilde W_0$, the solution consists of 
a BPS wall and an anti-BPS wall at fixed points 
$\tilde \phi=0$ and $\tilde \phi=\pi$ of the 
orbifold supplemented by a small 
positive cosmological constant at $\tilde \phi=0$ 
and a large negative cosmological constant 
at $\tilde \phi=\pi$. 
We can similarly construct solutions with arbitrary 
winding number with or without the orbifold symmetry 
by connecting solution curves at $\tilde \phi = n\pi$ 
with arbitrary integer $n$. 
These solutions may be regarded natural 
in the sense that fixed points 
occur where the scalar field is localized as walls. 

\begin{figure}[h]
\begin{center}
\includegraphics[width=15cm,height=4cm]
{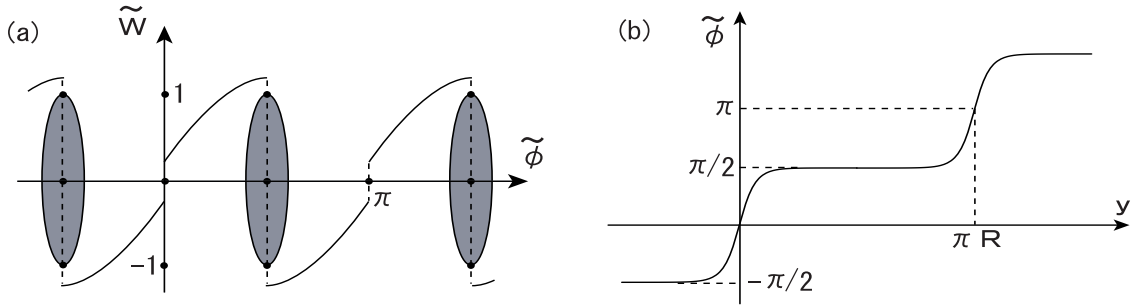}
 \caption{\small 
 Non-BPS solution with unit winding number with negative boundary 
 cosmological constant at $\tilde \phi=\pi/2$. 
 (a) $\tilde \phi \tilde W$ 
 plane, and (b) $\tilde \phi(\tilde y)$. 
 }
 \label{fig:half_wind_sol}
\end{center}
\end{figure}

If we allow the fixed points of orbifold at 
the mid point between two walls, 
we can connect solution curves at $\tilde \phi=n\pi + 
\pi/2$. 
{}For instance we can 
 switch solution curves at 
$\tilde \phi=\pi/2$ 
from 
$\tilde W(\tilde \phi;(0, \tilde W_0>0))$ to 
 $\tilde W(\tilde \phi; (\pi, -\tilde W_0))$ 
and then at $\tilde \phi=\pi$ 
to $\tilde W(\tilde \phi; (\pi, \tilde W_0))$. 
This configuration has a periodicity $\tilde \phi=\tilde \phi+\pi$ 
and has a symmetry of $S^1/(Z_2\times Z_2)$ orbifold as shown in 
Fig.\ref{fig:half_wind_sol}. 
In the limit of small $\tilde W_0$, this solution can be regarded 
as a model essentially replacing the positive 
cosmological constant by a smooth physical wall configuration 
of scalar field whereas the negative cosmological constant 
remains almost the same as in the original model \cite{RS}, \cite{RS2}. 
Actually we can also connect solution curves at arbitrary points 
in $\tilde \phi$ by switching to the curve with the same 
derivative at the same $\tilde \phi$. 

\begin{figure}[h]
\begin{center}
\includegraphics[width=15cm,height=4cm]
{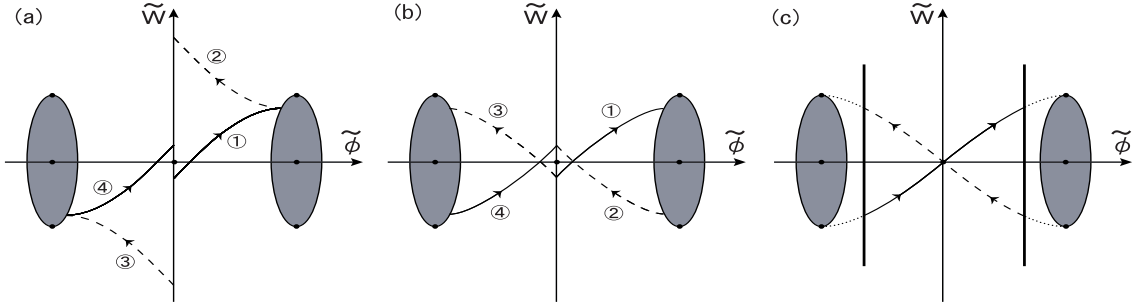}
 \caption{\small 
 Non-BPS solutions without winding number in $\tilde \phi \tilde W$ 
 plane. 
 Discontinuities in $\tilde W(\tilde \phi)$ at 
 $\tilde \phi=0, \pm\pi$ correspond to boundary cosmological 
 constants. 
 (a) Smooth $\tilde \phi$ with two types of 
 boundary cosmological constants 
 at $y=0, \pi R/2$. 
 (b) $\tilde \phi$ with only 
 one type of boundary cosmological constant 
 but with a break of scalar field at $y=\pi R/2$. 
 (c) Using a piece of the BPS solution and boundary source for $\phi$. 
 }
 \label{fig:no_wind_sol}
\end{center}
\end{figure}

Another possibility to connect different solution curves 
is to switch to curves with opposite sign $d\tilde W/d\tilde \phi$. 
This is possible only at the boundary of the forbidden region 
where $d\tilde W/d\tilde \phi=0$. 
If we connect solutions at the boundary of the 
forbidden region, we can do without 
putting boundary cosmological constant by 
switching to the curve starting from 
 the same value of $\tilde W$ with 
 the opposite sign for $d\tilde W/d\tilde \phi$. 
A typical example with no winding is shown 
in Fig.\ref{fig:no_wind_sol}(a).
This solution also has the orbifold symmetry $S^1/Z_2$. 

If we allow boundary cosmological constants at the boundary 
of the forbidden region, 
we can connect solutions 
by switching to the curve starting from 
 the $\tilde W$ as well as $d\tilde W/d\tilde \phi$ 
of opposite sign at the boundary of forbidden region. 
We illustrate the simplest of such solution 
in Fig.\ref{fig:no_wind_sol}(b). 
This solution has the symmetry $S^1/(Z_2\times Z_2)$ 
and has no winding number. 
If we relax our principle and allow 
the boundary action to contain a source term for 
the scalar field $\phi$, we have more varieties of solutions. 
{}For instance we can even use the BPS solution to 
construct a solution, simplest of which 
is illustrated in 
Fig.\ref{fig:no_wind_sol}(c).
It has the orbifold symmetry $S^1/Z_2$ but without any cosmological constant 
at one of the fixed point where the scalar field energy density is localized. 
This is a model replacing the positive cosmological 
constant of the original model \cite{RS} completely by a smooth thick 
wall configuration of scalar field at the cost of 
having a boundary source term for the scalar field in the other fixed 
point. 
We can also construct more complicated solutions with or without 
the orbifold symmetry by combining these 
various types of boundary cosmological constants.

The precise amount of the cosmological constants and the distance 
between the walls are determined by Eqs.(\ref{1st-W}) and 
(\ref{eq:bound_cod_A}). 
We will give a systematic analytic approximation valid for small gravitational 
coupling below.

\vspace{4mm}

{\large\bf 
A systematic approximation 
for weak gravity 
}

\vspace{2mm}

To 
establish the limit of no gravity ($\alpha \rightarrow 0$) 
for the first order differential equations (\ref{1st-W}) 
and (\ref{eq:DFGK_eq}), 
let us first examine 
the equation of motion for scalar field 
$\tilde \phi$ without gravity using the scalar potential 
 $\tilde V(\tilde \phi;{\alpha=0})$ in the absence of gravity 
\begin{equation}
{d\tilde V(\tilde \phi; {\alpha=0}) \over d\tilde{\phi}}
=
2{d^2\tilde{\phi} \over d\tilde y^2}. 
\end{equation}
By multiplying $d\tilde \phi/d\tilde y$, we can integrate 
it once with an integration constant $w_0$ 
\begin{equation}
\tilde V(\tilde \phi;{\alpha=0})=
\left({
d\tilde{\phi} \over 
d\tilde y}
\right)^2-3w_0^2. 
\end{equation}
This first order differential equation can be identified as 
the no gravity limit of Eq.(\ref{1st-W}), if a function 
$\tilde W(\tilde \phi)$ is defined by 
$d\tilde W/d\tilde \phi=d\tilde \phi/d\tilde y$. 
Because of Eq.(\ref{1st-W}), Eq.(\ref{eq:DFGK_eq}) dictates 
that the limit of vanishing gravitational coupling is given by 
\begin{equation}
\alpha \rightarrow 0, \qquad 
\alpha \tilde W({\tilde \phi}=0) \equiv w_0 \quad {\rm fixed}. 
\end{equation}
Eqs.(\ref{eq:DFGK_eq}) and (\ref{1st-W}) show that $w_0=0$ corresponds to 
the BPS solution. 

Since 
$\tilde W_{0}\equiv \tilde W({\tilde \phi}=0)\sim 1/\alpha\rightarrow \infty$, 
we define a function $\bar W$ more appropriate for 
an expansion in powers 
of gravitational coupling $\alpha$  
\begin{equation}
\bar W(\tilde \phi) 
\equiv 
{\tilde W}(\tilde \phi)-{w_0 \over \alpha} 
= \sum_{n=0}^{\infty} \bar W^{(n)}(\tilde \phi), 
\quad 
\tilde \phi(\tilde y) = \sum_{n=0}^{\infty} \tilde \phi^{(n)}(\tilde y),
\quad 
A(\tilde y) = \sum_{n=0}^{\infty} A^{(n)}(\tilde y).
\end{equation}
The first order differential equations appropriate for an expansion 
in powers of $\alpha$ is 
\begin{equation}
{d \bar W \over d \tilde \phi}
=
{d \tilde \phi \over d \tilde y}
=\pm \sqrt{\tilde V(\tilde \phi)
 + 3 \left(w_0 + \alpha {\bar W}(\tilde \phi) \right)^2 },
\quad 
{d A \over d \tilde y}
=-\alpha w_0 
-\alpha^2 \bar W(\tilde \phi). 
\end{equation}
It is interesting to note that the first correction is 
of order $\alpha\equiv \Lambda/(gM_{\rm P})$ rather than 
the usual $\alpha^2$. 
{}For our specific model of gravitationally corrected periodic potential, 
we obtain 
\begin{equation}
{d \bar W^{(0)} \over d \tilde \phi}
=
{d \tilde \phi^{(0)} \over d \tilde y}
=
\sqrt{
\cos^2 \tilde \phi + 3 w_0^2 }, 
\quad 
{d \bar W^{(1)} \over d \tilde \phi}
=
{d \tilde \phi^{(1)} \over d \tilde y}
=
 { 3\alpha w_0 {\bar W^{(0)}} \over \sqrt{
\cos^2 \tilde \phi + 3 w_0^2 }}, \cdots 
\label{eq:weak_gr}
\end{equation}
\begin{equation}
{d A^{(0)} \over d \tilde y}
=0, 
\quad 
{d A^{(1)} \over d \tilde y}
=
-\alpha w_0, 
\quad 
{d A^{(2)} \over d \tilde y}
=-\alpha^2 \bar W^{(0)}(\tilde \phi), \cdots.
\label{eq:weak_gr_A}
\end{equation}
{}The zero-th order equation for $\tilde \phi^{(0)}(y)$ 
in (\ref{eq:weak_gr}) gives 
precisely our exact non-BPS 
solution which was obtained in the case of global SUSY \cite{MSSS2} 
\begin{equation}
 \tilde \phi^{(0)}(\tilde y)
={\rm am}\left({\tilde y-\tilde y_0 \over k}, k\right), 
\qquad 
k\equiv {1 \over \sqrt{1+3w_0^2}}, 
\label{eq:global_lim_phi}
\end{equation}
\begin{equation}
\bar W^{(0)}(\tilde \phi)
=
{1 \over k}\int_0^{\tilde \phi}d\theta 
\sqrt{1-k^2\sin^2 \theta}
=
{1 \over k}E\left(\tilde \phi, k\right), 
\label{eq:global_lim_WA}
\end{equation}
where am is the amplitude function and $E$ is the elliptic 
integral of the second kind. 
The warp factor to the zero-th order is given by an irrelevant 
normalization $A^{(0)}(\tilde y)=$ 
constant. 

Using Eq.(\ref{eq:cosm_const_W}), we can now determine 
to the lowest order in weak 
gravitational coupling the boundary cosmological constant 
 needed to satisfy 
the boundary condition (\ref{eq:bound_cod_A}). 
{}For instance, the unit winding solution in Fig.\ref{fig:wind_sol} 
requires the boundary cosmological constants $\lambda_0$ 
at $\tilde \phi=0$ and $\lambda_1$ 
at $\tilde \phi=\pi$ 
\begin{equation}
\lambda_0=4{M_{\rm P}\Lambda^2 \over g}w_0, 
\qquad 
\lambda_1=-4{M_{\rm P}\Lambda^2 \over g}w_0 
- 8{\Lambda^3 \over g^2 k}E(k), 
\end{equation}
where $E(k)$ is the complete elliptic integral of the second kind 
$E(k)\equiv E(\pi/2, k)$. 
Similarly other situations can also be expressed in terms of 
elliptic integrals. 

A systematic approximation in powers of gravitational coupling 
is given by Eq.(\ref{eq:weak_gr}). 
{}For instance the first order correction is given by integrating 
the second of Eq.(\ref{eq:weak_gr})
\begin{equation}
\tilde y-\tilde y_0= \int_0^{\tilde \phi^{(1)}}d\theta
{\sqrt{1-k^2\sin^2 \theta} \over 3\alpha w_0 E(\theta, k)}, 
\end{equation}
\begin{equation}
\bar W^{(1)} (\tilde \phi)
=\int_0^{\tilde \phi} d\theta {3\alpha w_0 E(\theta, k) \over 
\sqrt{1-k^2\sin^2 \theta}}, 
\qquad 
A^{(1)}(\tilde y)=
-\alpha w_0 \tilde y
.
\end{equation}
We can obtain gravitational corrections to any desired order 
with increasing complexity. 
We stress that SUSY is not needed 
for our method to work 
similarly 
to Ref.\cite{DFGK}. 
Therefore it should be useful to obtain 
solutions for theories with or without gravity 
irrespective of presence or absence of SUSY. 

\begin{figure}[h]
\begin{center}
\includegraphics[width=7cm,height=4cm]
{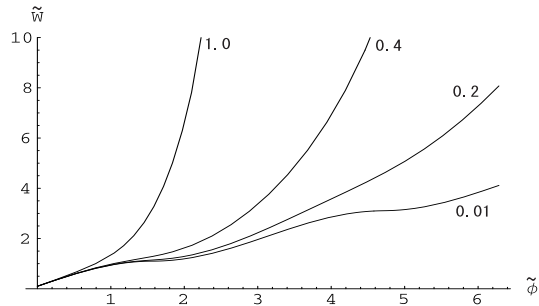}
 \caption{\small 
 Solution curves of Eq.(\ref{1st-W}) 
 in $\tilde \phi \tilde W$ plane 
 with larger gravitational coupling $\alpha=\Lambda/(gM_{\rm P})=
 0.01, \cdots, 1.0$. 
 The curve stays almost the same for $\alpha \le 0.01$. 
 }
 \label{fig:strong_gravity}
\end{center}
\end{figure}
Let us also examine the situation with stronger gravitational 
coupling. 
 The curve stays almost the same for $\alpha \le 0.01$. 
We see in Fig.\ref{fig:strong_gravity} that the larger values $\alpha > 0.4$ 
reveal an 
early emergence of singularities even for relatively smaller values 
of $\tilde \phi$. 
However, qualitative features remain quite similar for small 
values of $\tilde \phi$. 
Therefore we expect non-BPS multi-wall configurations are still 
possible for medium gravitational coupling. 
It is interesting to note that our BPS solution exists even with 
strong gravitational couplings at least classically. 

Finally we wish to note that the existence of the global SUSY limit 
(\ref{eq:global_lim_phi}), (\ref{eq:global_lim_WA}) implies that 
we are sure that there is a mass gap for fluctuation modes of 
the scalar fields. 
There is a possibility for our wall scalar field $\phi$ to serve as 
a Goldberger-Wise type stabilizer \cite{GoWi}. 
We will settle the issue of stability by analyzing 
the possible new modes associated with the gravity 
supermultiplet in a subsequent report. 


\begin{center}
{\bf Acknowledgements}
\end{center}
One of the authors (N.S.) thanks useful discussion with M.~Fukuma, 
 D.~Ida, T.~Kugo, N.~Sasakura, 
 T.~Shiromizu, M.~Siino, T.~Tanaka, Y.~Tanii, and A.~Van Proeyen. 
Two of the authors (N.M.,N.S.) are indebted to Y.~Sakamura and R.~Sugisaka 
for a collaboration in an early stage. 
This work is supported in part by Grant-in-Aid for Scientific 
Research from the Ministry of Education, Science and Culture 
 13640269. 
N.M.~is supported 
by the Japan Society for the Promotion of Science for Young Scientists 
(No.08557).



\begin{thebibliography}{100}
 \bibitem{DGSW} S. Dimopoulos and H. Georgi, 
               {\it Nucl.\ Phys.\ }~{\bf B193} (1981) 150; 
               N. Sakai, Z.\ f.\ Phys.\ {\bf C11} (1981) 153;
               E. Witten, {\it Nucl.\ Phys.\ }~{\bf B188} 
               (1981) 513;
               S.~Dimopoulos, S.~Raby, and F.~Wilczek, 
               {\it Phys.\ Rev.\ }~{\bf D24} (1981) 1681.
%
%
 \bibitem{LED}N.~Arkani-Hamed, S.~Dimopoulos and G.~Dvali, 
             {\em Phys. Lett.} {\bf B429} (1998) 263 
             [hep-ph/9803315]; 
             I.~Antoniadis, N.~Arkani-Hamed, S.~Dimopoulos 
             and G.~Dvali, 
             {\em Phys. Lett.} {\bf B436} (1998) 257 
             [hep-ph/9804398]. 
 \bibitem{RS}L.~Randall and R.~Sundrum, 
             {\em Phys. Rev. Lett.} 
             {\bf 83} (1999) 3370 [hep-ph/9905221].
 \bibitem{RS2}L.~Randall and R.~Sundrum, 
             {\em Phys. Rev. Lett.} {\bf 83} (1999) 4690 
             [hep-th/9906064].
 \bibitem{ABN}R.~Altendorfer, J.~Bagger, and D.~Nemeschansky, 
               {\it Phys.\ Rev.}~{\bf D63} (2001) 125025,
             [hep-th/0003117];
              T.~Gherghetta, A.~Pomarol, 
             {\it Nucl.\ Phys.}~{\bf B586} (2000) 141,  
             [hep-ph/0003129]; 
             A.~Falkowski, Z.~Lalak, and S.~Pokorski, 
             {\em Phys. Lett.}~{\bf B491} (2000) 172,
             [hep-th/0004093];
             E.Bergshoeff, R.~Kallosh, and A.~Van Proeyen, 
             {\em JHEP} {\bf 0010} (2000) 033, [hep-th/0007044].
 \bibitem{CGR}M.~Cvetic, S.~Griffies and S.~Rey,
             {\it Nucl.\ Phys.\ } {\bf B381} (1992) 301 
             [hep-th/9201007];
             M.~Cvetic, and H.H.~Soleng,
             {\em Phys. Rep.} {\bf B282} (1997) 159 
             [hep-ph/9804398]. 
 \bibitem{BCY}F.A.~Brito, M.~Cveti\u{c}, and S.C.~Yoon, 
             {\it Phys.\ Rev.}~{\bf D64} (2001) 064021,
             [hep-ph/0105010]. 
 \bibitem{DFGK}O.~DeWolfe, D.Z.~Freedman, S.S.~Gubser, and A.~Karch, 
              {\it Phys.\ Rev.\ }~{\bf D62} (2000) 046008, 
             [hep-th/9909134]. 
 \bibitem{SkTo}K.~Skenderis, and P.K.~Townsend, 
              {\em Phys. Lett.} {\bf B468} (1999) 46,
             [hep-th/9909070];
             G.W.Gibbons and N.D.~Lambert, 
             {\em Phys. Lett.} {\bf B488} (2000) 90,
             [hep-th/0003197]. 
 \bibitem{GiLa}
             J. Garriga, and T. Tanaka, 
             {\em Phys. Rev. Lett.} {\bf 84} (2000) 2778,  
             [hep-th/9911055];
              Csaba Csaki, J.~Erlich, T.~J.~Hollowood, and Y.~Shirman, 
             {\it Nucl.\ Phys.\ } {\bf B581} (2000) 309, 
            [hep-th/0001033];
             S.~Ichinose, 
             {\it Phys.\ Rev.\ }~{\bf D65} (2002) 084038l,
             [hep-th/0008245];
             K.~Behrndt, C.~Herrmann, J.~Louis, and S.~Thomas, 
             {\em JHEP} {\bf 01} (2001) 011, [hep-th/0008112];
             M.~Duff, J.T.~L\"u, and C.~Pope, 
             {\it Nucl.\ Phys.\ } {\bf B605} (2001) 234 
             [hep-th/0009212];             
             A.~Ceresole, G.~Dall'Agata, R.~Kallosh, and A.~Van Preoyen, 
             {\it Phys.\ Rev.\ }~{\bf D64} (2001) 104006,
             [hep-th/0104056];
             M.~Gremm, 
              {\it Phys.\ Rev.\ }~{\bf D62} (2000) 044017, 
             [hep-th/0002040];
             M.~Cvetic, and N.D.~Lambert, 
             {\em Phys. Lett.} {\bf B540} (2002) 301, 
             [hep-th/0205247].
 \bibitem{BeDA}K.~Behrndt, and G.~Dall'Agata, 
             {\it Nucl.\ Phys.\ } {\bf B627} (2002) 357, 
             [hep-th/0112136].
 \bibitem{Sasakura}N.~Sasakura, JHEP 0202 (2002) 026
             {\em JHEP} {\bf 0202} (2002) 026, [hep-th/0201130].
 \bibitem{WittenOlive} E.~Witten and D.~Olive, 
             {\it Phys.\ Lett.\ } {\bf B78} (1978) 97.
 \bibitem{DvSh}G.~Dvali and M.~Shifman, 
             {\em Phys. Lett.} {\bf B396} 
             (1997) 64 [hep-th/9612128]; 
            A.~Kovner, M.~Shifman, and A.~Smilga, 
            {\it Phys.\ Rev.\ } {\bf D56} (1997) 7978 
            [hep-th/9706089]; 
            A.~Smilga and A.~Veselov, 
            {\it Phys.\ Rev.\ Lett.\ } {\bf 79} (1997) 4529 
            [hep-th/9706217]; 
            B.~Chibisov and M.~Shifman, 
            {\it Phys.\ Rev.\ } {\bf D56} (1997) 7990, 
            [hep-th/9706141];
            J.~Edelstein, M.L.~Trobo, F.~Brito and D.~Bazeia, 
            {\it Phys.\ Rev.\ } {\bf D57} (1998) 7561 
            [hep-th/9707016];
             V.~Kaplunovsky, J.~Sonnenschein, and 
             S.~Yankielowicz, 
             {\it Nucl.\ Phys.\ } {\bf B552} (1999) 209 
             [hep-th/9811195];
             B.~de Carlos and J.~M.~Moreno,
            {\it Phys.\ Rev.\ Lett.\ } {\bf 83} (1999) 2120 
            [hep-th/9905165];
            D.~Binosi and T.~ter Veldhuis, Phys. Rev. D 63: 085016 (2001), 
            [hep-th/0011113].
%
\bibitem{OINS}H. Oda, K. Ito, M. Naganuma and N. Sakai, 
              {\em Phys. Lett.} {\bf B471} (1999) 148 
              [hep-th/9910095]; 
              K. Ito, M. Naganuma, H. Oda and N. Sakai, 
              {\em Nucl. Phys.} {\bf B586} (2000) 231 
              [hep-th/0004188]; 
             {\em Nucl. Phys. Proc. Suppl.} {101} (2001) 304 
               [hep-th/0012182].
 \bibitem{MSSS} N.~Maru, N.~Sakai, Y.~Sakamura, and 
                R.~Sugisaka, {\it Phys. Lett.} {\bf B496} 
                (2000) 98, [hep-th/0009023]. 
 \bibitem{MSSS2} N.~Maru, N.~Sakai, Y.~Sakamura, and R.~Sugisaka, 
                {\it Nucl.~Phys.} {\bf B616} 
                (2001) 47,  [hep-th/0107204];
                ``SUSY Breaking by Stable non-BPS Walls'', 
                  in the Proceedings of the 10th Tohwa 
                international symposium on string theory [hep-th/0109087];
                ``SUSY Breaking by stable non-BPS configurations'', 
                to appear in the Proceedings of Corfu International Summer 
                Institute, [hep-th/0112244].
 \bibitem{SaSu} N.~Sakai, and R.~Sugisaka, 
                ``Winding number and non-BPS bound states of walls 
                in nonlinear sigma models'' to appear in {\it Phys.\ Rev.\ }, 
                [hep-th/0203142]. 
%
\bibitem{WessBagger} J.~Wess and J.~Bagger, 
                 ``Supersymmetry and Supergravity'', 1991, Princeton 
                  University Press. 
\bibitem{GoWi}W.D.~Goldberger and M.B.~Wise, 
            {\it Phys.\ Rev.\ Lett.\ } {\bf 83} (1999) 4922 
            [hep-th/9907447].
%
%
%
\end{thebibliography}
\end{document}